# May silicene exist?

E.F.Sheka

*Peoples' Friendship University of the Russian Federation, General Physics Department, Laboratory of Computational Nanotechnology, 117198 Moscow, Russia*

e-mail: sheka@icp.ac.ru

**Abstract**. The letter presents arguments, supported by quantum-chemical calculations, against silicene to be produced

Silicon-based nanotechnology has attracted a hottest attention since the first steps of nanotechnology to become a reality. Silicon-based nanoelectronics seemed an obvious extension of conventional silicon microtechnology. The first peak of interest was connected with the adventure of STM "atom writing" on the silicon crystal surfaces. A controllable deposition and/or extraction of silicon atoms from the surfaces seemed to open a direct way to design electronic nanochips of any kind. However, the works, pioneered by Prof.M.Aono and his team in Japan as well as by many other groups through over the world, met serious difficulties on the way that showed unexpected complications connected with the surfaces properties. Thus, the most promising Si(111)(7x7) surface occurred to be metallic and magnetic in contrast to semiconductive and non-magnetic bulk silicon. Silicon nanoelectronics has not met with success at that time, although undertaken efforts have stimulated a large realm of silicon surface science and the current extended materials nanoarchitechtonics (see projects of the International Center for Materials Nanoarchitectonics at NIMS in Japan).

The next pulse of interest is observing nowadays. It has been stimulated by extreme expectations connected with graphenium nanoprocessors. However, despite the reigning optimism about the devices, the graphene discoverers pointed out that the processors appearance is unlikely for the next 20 years [1] since replacement of the current silicon electronics technology is a tough hurdle. And again a compatibility of silicon-based nanoelectronics with the conventional one has stressed attention to the question if carboneous graphene can be substituted by its siliceous counterpart named as silicene. Meeting the demands, the December internet news has brought information on "epitaxial growth of graphene-like silicon nanoribbons" [2]. The report, based on the hexagon-patterned accommodation of silicon atoms adsorbed on the [110] Ag surface, has heralded the silicene manifestation and is full of exciting applications to be expected.

However, under detailed examination, the situation does not seem so transparent and promising. To clear the said let us specify basic terms. First, we have to improve what is implied under the term *silicene*. If any hexagon-packed structure of silicon atoms can be named silicene, then it has been known for long ago as crystalline structure seen from the (111) surface as well as many other things such as widely known silicon nanowires. As well known, in all these cases, four valence electrons of each silicon atoms form $sp^3$ configuration and nobody pretended to look for a similarity between these species and carboneous graphene. Therefore, not hexagon packing itself but a mono-atom-thick hexagon structure that dictates $sp^2$ configuration for atom valence electrons meets requirements of comparison of silicene to graphene. Obviously, similar hexagon patterns should form the ground for silicon nanotubes (SiNTs). Only under these conditions graphene and silicene, as well as carbon nanotubes (CNTs) and SiNts can be considered on the same basis.

As for theoretical consideration, performed computations of silicene [3] and SiNTs [4-6] meet the requirement completely. Oppositely, experimental reports are full of appealing to SiNTs (see a brief review [7]) and silicene [1] (in the first announcement of the finding observed [8], the latter was attributed to silicon nanowires) in spite of evident $sp^3$ configuration of silicon atoms in the structures observed. The fact has been accepted by the experimentalists themselves. But a temptation to disclose SiNts and silicene is so strong that the difference in the electron configuration is simply omitted. A detailed analysis of the available experimental data shows that silicon structures that can be compared to CNTs and graphene have not been observed. If we remember that fullerene $Si_{60}$ has not been produced as well we have to accept the availability of a serious reason that causes so drastic difference between carboneous and siliceous analogues.

The problem is not new with respect to nanotechnology and is rooted deeply so that "...A comparison of the chemistry of tetravalent carbon and silicon reveals such gross differences that the pitfalls of casual analogies should be apparent" [9]. Enough to remind that there is no either silicoethylene or silicobenzene as well as other aromatic molecules. A widely spread standard statement that "silicon does not like $sp^2$ configuration" just postulates the fact but does not explain the reason of such behavior. The reason has been disclosed for the first time when answering question why fullerene $Si_{60}$ does not exist [10]. The answer addresses changing in electron interaction for the two species when their electron configurations are transformed from $sp^3$ to $sp^2$-type. The interaction of two odd electrons, which are formed under transformation at any interatomic bond, depends on the corresponding distance $R_{int}$ that is ~1.5 times larger for Si-Si chemical bonds with respect to C-C ones due to larger size of the atom. As was shown, generally, the distance $R_{int} = 1.4 A$ is critical for these electrons to be covalently coupled [11]. Above the distance the electrons become effectively unpaired, therewith the more the larger the distance. C-C bond length of benzene just coincides with the limit that provides a complete covalent bonding of the electrons, transforming them into widely known $\pi$ electrons. But in fullerenes (both $C_{60}$ and $C_{70}$), as well as in CNTs and graphene the distances fill intervals of 1.38-1.47, 1.39-1.46, and 1.39-1.43A, respectively. Evidently, some of bonds exceed the limit value that causes partial exclusion of odd electrons from the covalent coupling which makes the molecular species partially radicals [10-13]. However, the radicalization is rather weak since only 10-15% of all odd electrons (equal to the number of atoms $N$) are unpaired. Oppositely to the case, $R_{int}$ in siliceous species is of 2.3-2.4A that causes a considerable unpairing of all odd electrons, up to the complete one, so that all siliceous species with expected $sp^2$ configuration should be many-fold radicals. Table 1 lists calculation results of the total number of unpaired electrons $N_D$ as well as a set of energetic parameters for a number of siliceous $sp^2$-configured species shown in Figure 1. The exploitation of unrestricted broken-spin symmetry Hartree-Fock (UBS HF) approach is not conceptually critical. Similar results should be expected from UBS

DFT as well, whilst different numerically due to different algorithms of taking electron correlation effects into account (see details in [13]). At the same time, a strong weakening of the odd electron interaction in siliceous species, that is pointed out in papers [3-6] as well, makes the use of either complete configurational interaction schemes or approximated UBS approaches absolutely mandatory.

A convincing confirmation of the said above follows from the data presented in Table 1. As seen from the table, there is a drastic lowering of the total energy of the species, constituting about 20-30% of the largest values, when close-shell restricted HF (RHF) scheme is substituted by open-shell UBS HF. This is an undisputable evidence of the weakness of electron interaction that causes spin-mixed character of both RHF and UBS HF solutions. Well theoretically grounded procedure, suggested by Noodelman [14], allows for determining the energy of pure spin states on the basis of UBS HF solution. Thus obtained pure-spin energies of the singlet state of the species are given in the table. As should be expected, the energy is lower than both RHF and UBS HF ones, while rather close to the latter. Another important advantage of UBS HF approach is its ability to determine the total number of effectively unpaired electrons $N_D$ [11]. The values presented in the table coincide quite well with the total numbers of silicon atoms in all cases ($N$) and exceed the values by the number of two-neighbor atoms ($N_2$) when hydrogen terminators are taken off from either of tubes ends or silicene edges. The finding exhibits that both silicon fullerene as well as SiNTs and silicene are many-fold radicals and cannot exist under ambient conditions. Optimism expressed in theoretical papers where fullerene $Si_{60}$ [15], SiNts [4-6], and silicene [3] were considered, is mainly due to that the calculations were performed in the close-shell approximation (similar to RHF) so that the problem concerning weak interacting odd electrons was not taken into account.

The odd-electron concept explains the failure of a controllable STM atom writing on the silicon surfaces as well [16]. Dangling bonds of the surface atoms are just another appearance of odd electrons but for $sp^3$-configured molecular configurations. A drastic weakening of these electrons interaction due to large distance between the corresponding atoms (adatoms, restatoms and so forth) forces the Si(111)(7x7)surface to behave as a space-extended many-fold radical. This explains both electron conductivity and magnetism of the surface due to creation of nearly degenerated states in the vicinity of the Fermi level as well as high delocalization of the odd electrons over the surface that prevents from a local fixation of either deposited or extracted atom which might provide a skeleton of a wished electronic nanochip. Radical character of the surfaces explains as well their extremely high chemical reactivity towards, say, oxidation that leads to a burning match process behaving like a propagation flame front [17]. A similar burning oxidation has been observed for silicon nanowires [18] that were called later [2] as silicene strips.

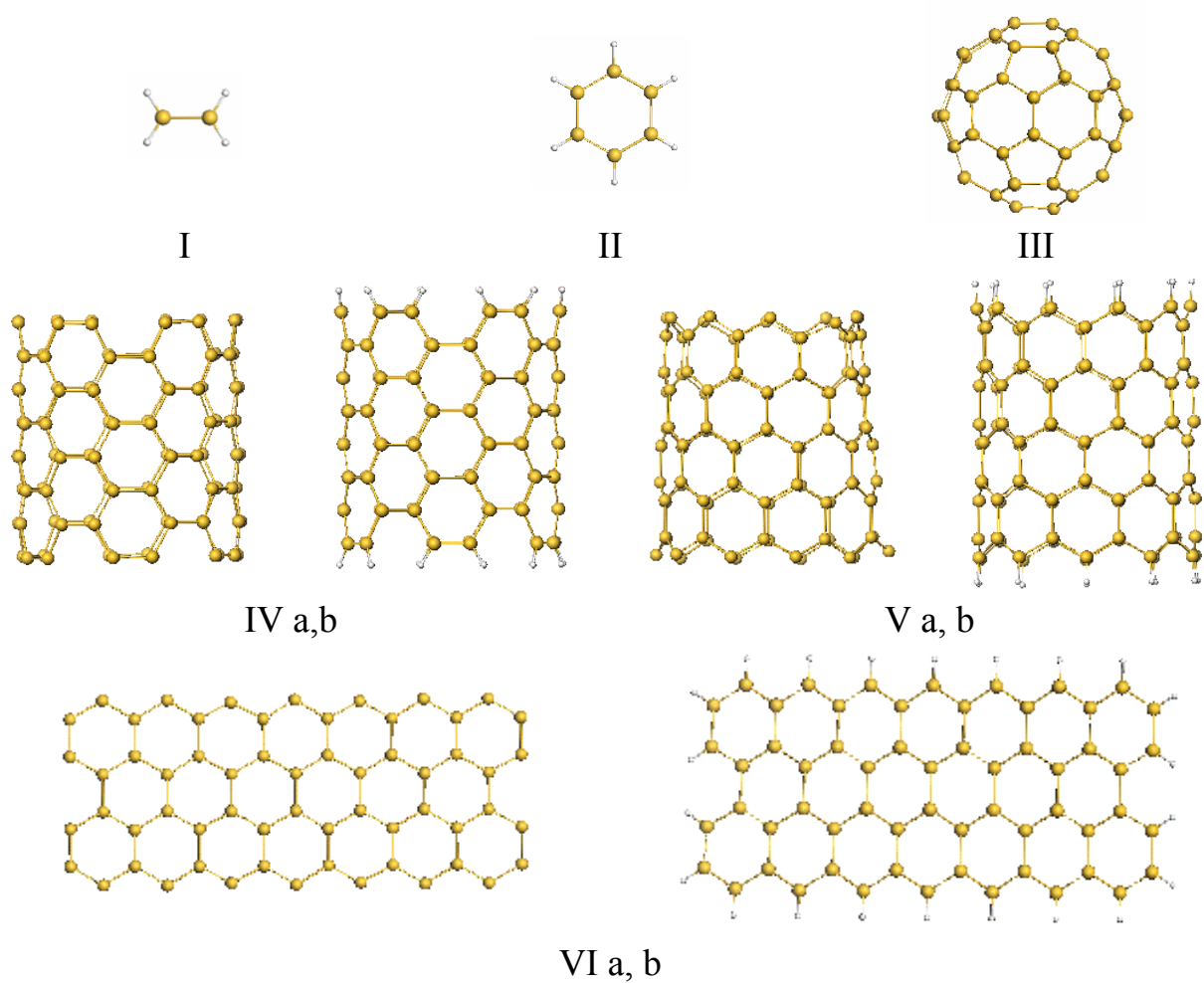

Figure 1. Equilibrated structures of *sp²*-configured siliceous species, UBS HF, singlet state: I- silicoethylene; II- silicobenzene; III- silicofullerene $Si_{60}$; IV- fragments of (6,6) SiNT with empty (a) and hydrogen-terminated (b) end atoms; V- the same but for (10,0) SiNT; VI- (7,3) silicene sheet with empty (a) and hydrogen-terminated (b) edges.

**Table 1**. Heats of formation, *kcal/mol* and the number of effectively unpaired electrons in $sp^2$-configured siliceous species (see Fig.1.)

| Species | $N(N_2)$ | $\Delta H_{RHF}$ | $\Delta H_{UHF}$ | $\Delta H (S=0)$ | $N_D$ |
|---|---|---|---|---|---|
| **I** | 2 | 54,502 | 48,949 | 39,021 | 0.88 |
| **II** | 6 | 144,509 | 121,246 | 108,6703 | 2.68 |
| **III** | 60 | 1295,988 | 1013,301 | 996,6374 | 62.48 |
| **IV a** | 96 (24)[1] | 2530,186 | 1770,909 | 1749,563 | 128 |
| **IV b** | 96 | 1943,14 | 1527,767 | 1505,48 | 95,7 |
| **V a** | 100 (20)[1] | 2827,727 | 1973,667 | 1958,54 | 115,05 |
| **V b** | 100 | 2119,596 | 1580,766 | 1559,64 | 100,12 |
| **VI a** | 60 (22)[1] | 1950,198 | 1359,437 | 1346,68 | 75,7 |
| **VI b** | 60 | 1253,39 | 1001,269 | 972,122 | 54,043 |

[1] The number of two-neighbor silicon atoms